\documentclass[11pt,a4paper]{article}

\usepackage[utf8]{inputenc}
\usepackage[T1]{fontenc}
\usepackage{amsmath,amssymb}
\usepackage{graphicx}
\usepackage{booktabs}
\usepackage{hyperref}
\usepackage{url}
\usepackage[numbers,sort&compress]{natbib}
\usepackage{enumitem}
\usepackage{xcolor}
\usepackage{listings}
\usepackage[margin=1in]{geometry}
\usepackage[protrusion=true,expansion=false]{microtype}
\hypersetup{
  colorlinks=true,
  linkcolor=blue,
  citecolor=blue,
  urlcolor=blue
}

\title{\textbf{Any2Speech: Borderless Long Speech Synthesis}}

\author{
  Xingchen Song$^{1,3}$\thanks{Core contributors.} \quad
  Di Wu$^{1,3}$\footnotemark[1] \quad
  Dinghao Zhou$^{1,3}$\footnotemark[1] \quad
  Pengyu Cheng$^{1}$\footnotemark[1] \\
  Hongwu Ding$^{1}$ \quad
  Yunchao He$^{1}$ \quad
  Jie Wang$^{1}$ \quad
  Shengfan Shen$^{1,2}$ \\
  Sixiang Lv$^{1,2}$ \quad
  Lichun Fan$^{1}$ \quad
  Hang Su$^{1}$ \quad
  Yifeng Wang$^{1}$ \quad
  Shuai Wang$^{2,3}$ \\
  Meng Meng$^{1}$ \quad
  Jian Luan$^{1}$ \\[6pt]
  $^{1}$ MiLM Plus, Xiaomi Inc., China \\
  $^{2}$ Nanjing University, China \\
  $^{3}$ WeNet Open Source Community
}

\date{}

\begin{document}

\maketitle

\begin{abstract}
Most existing text-to-speech (TTS) systems either synthesize speech sentence by sentence and stitch the results together, or drive synthesis from plain-text dialogues alone. Both approaches leave models with little understanding of global context or paralinguistic cues, making it hard to capture real-world phenomena such as multi-speaker interactions (interruptions, overlapping speech), evolving emotional arcs, and varied acoustic environments. We introduce the Borderless Long Speech Synthesis framework for agent-centric, borderless long audio synthesis. Rather than targeting a single narrow task, the system is designed as a unified capability set spanning VoiceDesigner, multi-speaker synthesis, Instruct TTS, and long-form text synthesis. On the data side, we propose a ``Labeling over filtering/cleaning'' strategy and design a top-down, multi-level annotation schema we call \textit{Global-Sentence-Token}. On the model side, we adopt a backbone with a continuous tokenizer and add Chain-of-Thought (CoT) reasoning together with Dimension Dropout, both of which markedly improve instruction following under complex conditions. We further show that the system is \textbf{Native Agentic} by design: the hierarchical annotation doubles as a Structured Semantic Interface between the LLM Agent and the synthesis engine, creating a layered control protocol stack that spans from scene semantics down to phonetic detail. Text thereby becomes an information-complete, wide-band control channel, enabling a front-end LLM to convert inputs of any modality into structured generation commands --- extending the paradigm from Text2Speech to borderless long speech synthesis.
\end{abstract}

\section{Introduction}
\label{sec:intro}

Human speech is far more than words read aloud. It conveys emotion, atmosphere, and interaction --- nuances that successive generations of TTS systems have tried to capture:

\textbf{Generation~1: Concatenative and Parametric Synthesis.}
Unit Selection~\citep{hunt1996unit} and HMM-based statistical methods~\citep{zen2009statistical} made it possible, for the first time, to turn arbitrary text into intelligible speech. Prosody remained stiff and quality mechanical, but the foundational problem was solved.

\textbf{Generation~2: End-to-End Neural TTS.}
Tacotron~\citep{tacotron}, WaveNet~\citep{wavenet}, FastSpeech~\citep{fastspeech}, VITS~\citep{vits}, and their successors brought a step change in naturalness and fidelity. Yet these models are still ``reading machines'' --- they faithfully vocalize whatever text they receive, with no explicit levers for emotion or rhythm.

\textbf{Generation~3: LLM-based TTS and Instruct TTS.}
VALL-E~\citep{valle}, CosyVoice~\citep{cosyvoice}, ChatTTS~\citep{chattts}, FishAudio~\citep{fishaudio}, and similar systems introduced LLM-grade capabilities and natural-language instruction interfaces. Users gained the ability to steer emotion, speed, and tone --- but only one sentence at a time.

Two fundamental gaps remain.

\textbf{Gap~1: Sentence-level control $\neq$ global coherence.}
Real speech rarely lives in isolation. In Poe's \textit{The Tell-Tale Heart}, the narrator begins with forced calm, is driven sentence by sentence toward collapse by the phantom heartbeat, and finally erupts in the scream ``louder! louder! louder!'' The arc from suppression to hysteria unfolds over dozens of sentences; how tense any one of them should be depends on how much tension has already built up and how close the breaking point is. An Instruct TTS system can make a single line sound ``tense,'' but it has no way of knowing \textit{how} tense it should be in context. Sentence-level synthesis and concatenation severs this internal continuity.

\textbf{Gap~2: Voice control $\neq$ scene completeness.}
Existing systems operate only on the voice itself. They cannot model the caf\'{e} chatter behind a casual conversation, the reverb of a theater stage, or the roar of a stadium crowd. Yet speech never occurs in a vacuum --- the acoustic scene is an integral part of the expression. The root cause is that plain-text dialogue representations (\texttt{<speaker1>text<speaker2>text}) strip away scene context, speaker profiles, paralinguistic cues, acoustic environments, and interaction dynamics.

We believe TTS is now entering \textbf{Generation~4: Native Agentic TTS}, defined by two shifts: from per-sentence control to \textbf{long-context global coherence}, and from voice-only generation to \textbf{full acoustic scene modeling}. Our system targets both.

Within this framing, the system should be understood not as a single-point feature, but as a unified multi-task capability set spanning VoiceDesigner, multi-speaker synthesis, Instruct TTS, and long-form text synthesis.

Concretely, this paper makes three contributions:

\begin{enumerate}[leftmargin=*]
  \item A new data paradigm for borderless long audio --- ``Labeling over filtering \& cleaning'' --- paired with a top-down \textit{Global-Sentence-Token} annotation schema (Section~\ref{sec:data}).
  \item Two training strategies --- CoT and Dimension Dropout --- validated on a continuous-tokenizer backbone, which substantially improve instruction following and expressiveness in complex scenarios (Section~\ref{sec:model}).
  \item A Native Agentic architecture in which the \textit{Global-Sentence-Token} schema serves as a structured semantic interface, turning the traditional narrow-band text channel into an information-complete wide-band control surface and enabling the leap from Text2Speech to borderless long speech synthesis (Section~\ref{sec:agentic}).
\end{enumerate}

\section{Data Strategy: Context is All You Need}
\label{sec:data}

Long audio demands context. Conventional annotations --- speaker IDs plus transcripts --- are nowhere near enough to reconstruct a complex scene. Dialogue context, mood, ambient sound, and emotion all need to be modeled jointly.

\subsection{Labeling over Filtering and Cleaning}
\label{sec:labeling}

Standard data pipelines chase purity: segments are aggressively discarded based on DNSMOS~\citep{dnsmos} scores, WER thresholds, or single-speaker checks. As documented by TouchTTS~\citep{touchtts}, such aggressive filtering typically retains only 10--30\% of the original corpus, severely constraining data scale. Worse, the most expressive data disappears along with the ``noise.''

We take the opposite stance: \textbf{label, don't filter.} Our analysis shows that high-expressiveness samples correlate strongly with acoustic messiness --- heated arguments, emotional multi-speaker exchanges, and overlapping talk are precisely the richest sources of vocal expressiveness, yet they are the first to be thrown out. Removing them systematically biases the training distribution away from real-world speech. We therefore keep overlaps, interruptions, interjections, and complex layering of voice with music and background events (movie commentary, sports broadcasts, etc.), preserving as much of the original data distribution as possible.

We go one step further: \textbf{label, don't clean.} Rather than denoising backgrounds, we model background sound, environmental noise, and voice as a unified whole, using fine-grained hierarchical annotation to characterize the acoustic scene. The result is data utilization above 90\%.

An obvious concern: does training on ``dirty'' data sacrifice controllability? Not when the annotations are detailed enough. Because the model learns explicit associations between background conditions and their textual descriptions, users can steer the acoustic environment at inference time --- ``quiet background'' for clean output, ``children crying in the background'' for a domestic scene, ``soft piano music'' for ambiance. The dirt is not noise; it is a new, controllable dimension of expression.

We also find an unexpected benefit: models trained exclusively on clean, simple data are prone to hallucination at inference time. Exposure to the full spectrum --- clean \textit{and} noisy, simple \textit{and} complex --- noticeably improves generation quality. We suspect this diversity will prove valuable for reinforcement learning as well, providing more discriminative reward signals, though systematic verification is left to future work.

\subsection{Global-Sentence-Token Hierarchical Annotation Schema}
\label{sec:schema}

To harness the complexity we have preserved, we impose a top-down annotation structure with three levels:

\begin{itemize}[leftmargin=*]
  \item \textbf{Global Layer} --- scene-level metadata: show format, style tags, speaker profiles (gender, age, vocal personality), overall emotional tone and trajectory, acoustic environment, and sound events.
  \item \textbf{Sentence Layer} --- per-utterance controls: tone, intonation, speed, volume, speaking intent, and background state, plus intra-sentence rich-text marks (interruption cues, tonal pivots, etc.).
  \item \textbf{Token Layer} --- phoneme-level precision: stress placement, polyphone disambiguation, connected-speech rules, and interjection durations.
\end{itemize}

The design mirrors professional voice acting. Before a recording session, actors receive detailed direction --- character age, personality, vocal texture (Global); scene context, emotional baseline, emotional arc (Global); per-line tone, pacing, dramatic pauses (Sentence); word stress, syllable stretch, breathing cues (Token). Our schema formalizes this workflow into a machine-readable protocol so that the synthesis engine receives the same multi-dimensional guidance a voice director would give.

A deliberate choice: \textbf{every dimension at every level is expressed as a free-form natural-language caption}, not a fixed enum or numeric code. Emotion becomes ``the tone gradually shifts from calm to agitated, peaking at the third sentence'' instead of \texttt{emotion=happy}; acoustic scene becomes ``faint caf\'{e} chatter with clinking cups'' instead of \texttt{noise\_level=0.3}. This matters because real emotion is rarely one-dimensional --- a tearful laugh, an excited whisper, a sarcastic compliment, a speech where forced calm barely masks rising panic --- these compound states resist discrete labels but yield readily to natural language. The caption format also aligns naturally with LLM input/output spaces, setting the stage for seamless agent integration.

\section{Model Architecture and Innovations}
\label{sec:model}

Rich, multi-dimensional ``dirty'' data demands a capable backbone. After controlled comparisons on the same training set, we chose a backbone with a \textbf{continuous tokenizer}, which we found essential for borderless generation. On top of this base, we introduce two strategies.

\subsection{Chain-of-Thought: Understand First, Synthesize Second}
\label{sec:cot}

Conventional TTS collapses understanding and vocalization into a single pass: the model jumps straight from text to audio, burying every prosody decision inside the network where it can be neither observed nor corrected. We contend that good synthesis, like good acting, requires a planning stage --- \textbf{comprehend the context, design the delivery, then speak}.

This philosophy resonates with three core language-intelligence capabilities: semantic analysis (choosing the right tone for the content), sentiment modeling (shaping the expressive strategy), and pragmatic reasoning (delivering identical words very differently depending on context). Traditional TTS folds all three into one opaque mapping; we make them explicit.

In practice, we partition the \textit{Global-Sentence-Token} annotations into two functional streams:

\begin{itemize}[leftmargin=*]
  \item \textbf{Instruct} --- user-supplied hard constraints: scene metadata (format, style, topic), speaker identity (gender, age, vocal qualities), and acoustic-environment ratings.
  \item \textbf{Think} --- the model's own expressive plan: global atmosphere and emotional arc, per-sentence tone, intonation, pace, volume, intent, and background state, plus phoneme-level pronunciation details (stress, liaison, tone sandhi).
\end{itemize}

At inference, the model does not jump to audio immediately. It first enters a planning phase, reading the global instructions, processing the text, and reasoning out the Think dimensions sentence by sentence. For a given utterance, it determines the tonal baseline, intonation contour, pace, communicative intent, and phonetic realization --- all as explicit, inspectable output. Only then does it synthesize. This two-stage ``think, then speak'' pipeline turns the black-box prosody decision into a traceable, interpretable, and editable reasoning chain, yielding noticeably better precision and global coherence in complex scenarios.

\subsection{Dimension Dropout}
\label{sec:dropout}

During training, we randomly mask certain Think dimensions (e.g., acoustic-environment descriptions or emotional trajectories). Masked slots remain empty --- the model is not asked to reconstruct them; it simply learns to produce high-quality audio with incomplete information. This prevents over-reliance on any single cue and delivers two practical benefits: (1)~the model follows the \textit{remaining} instructions more faithfully when some are missing, and (2)~users at inference time can specify only the dimensions they care about and still get good results.

\section{Native Agentic Architecture}
\label{sec:agentic}

The system described so far is, at its core, a Text2Speech engine that accepts richly annotated text and produces complex audio. We claim it is also \textbf{Native Agentic} --- but not in the trivial sense that ``it takes text in, so you can put an LLM in front.'' By that criterion every TTS system qualifies, and the term loses meaning.

The real distinction is \textbf{interface bandwidth}. A conventional TTS text interface is narrow-band: it carries lexical content and perhaps a speaker ID or a style tag. Even if the upstream LLM understands scene, emotion, and acoustics, all of that understanding is crushed to fit through the narrow pipe; only the words survive.

In our system, the \textit{Global-Sentence-Token} schema widens the pipe into a \textbf{full-bandwidth, information-complete} control channel. Scene positioning, speaker profiles, emotional arcs, acoustic environments, and interaction dynamics all travel alongside the text. Because the model has been trained to map these multi-dimensional descriptions to audio, a front-end LLM that restructures its understanding into the schema's format can have that understanding faithfully executed by the synthesis engine --- a stark contrast to the information loss inherent in traditional architectures.

This wide-band interface also unlocks modality extension. Text is a universal cross-modal bridge: a front-end LLM can ingest plain instructions, long videos, or any other input, distill its understanding into \textit{Global-Sentence-Token} commands, and hand them to the synthesis engine for end-to-end audio generation. The result is a shift from Text2Speech to borderless long speech synthesis.

\subsection{Structured Semantic Interface}
\label{sec:interface}

At a higher level of abstraction, the \textit{Global-Sentence-Token} schema functions as a \textbf{Structured Semantic Interface} --- a standardized contract between the LLM Agent and the synthesis engine. Its three layers map onto a \textbf{Layered Control Protocol Stack}: Global as the session/application layer (scene context, speaker profiles, emotional arcs), Sentence as the transport layer (per-utterance tone, intent, prosody), and Token as the physical layer (phoneme-level acoustic detail).

The LLM translates natural-language intent into control signals at each layer; the engine executes them. Neither the user nor the agent needs to know how audio is actually synthesized --- the interface abstracts it away while still allowing fine-grained intervention, from high-level mood down to individual phoneme shaping.

\subsection{Context Efficiency Advantage}
\label{sec:context}

Compared with end-to-end spoken-dialogue models, the Agent architecture enjoys a significant \textbf{context efficiency edge}. End-to-end models must feed the full conversation history into the network, incurring token costs that grow linearly with each turn. Worse, attention's well-documented long-range decay (the ``lost in the middle'' phenomenon~\citep{lostinmiddle}) means that earlier context is progressively ignored --- an unconscious, uncontrollable lossy compression.

The Agent architecture replaces brute-force context injection with \textbf{deliberate, schema-guided compression}. The LLM distills conversation history into a compact emotion-and-context state, mapped onto \textit{Global-Sentence-Token} dimensions. The synthesis engine receives not raw history but a semantically distilled instruction set. Cross-turn emotional coherence is preserved, while the engine's compute is freed to focus on the present utterance rather than on recalling the past.

\section{Experiments and Evaluation}
\label{sec:exp}

Evaluating borderless long audio synthesis is fundamentally challenging: the relevant quality dimensions --- emotional-arc coherence, multi-speaker interaction naturalness, acoustic-scene fidelity, instruction-following accuracy --- are deeply intertwined, and no existing automated metric covers them adequately.

We explored several candidates: CLAP~\citep{clap} for audio-text alignment, Audio Captioning models for reverse-checking semantic consistency, and classic signal-level metrics (DNSMOS~\citep{dnsmos}, PESQ~\citep{pesq}, POLQA~\citep{polqa}). None proved satisfactory:

\begin{itemize}[leftmargin=*]
  \item \textbf{CLAP and similar cross-modal models}, trained predominantly on short clips, lack the discriminative power needed for long-audio phenomena like sustained emotional arcs and multi-speaker dynamics.
  \item \textbf{Audio Captioning models} operate at a description granularity far too coarse to cover the rich semantics encoded in a three-layer \textit{Global-Sentence-Token} annotation; paralinguistic subtleties (tonal pivots, interruption timing, emotional gradients) largely escape them.
  \item \textbf{Signal-level metrics} (DNSMOS, PESQ, POLQA) measure acoustic quality at the waveform level and are blind to scene modeling, expressiveness, and instruction adherence.
\end{itemize}

We therefore cannot present quantified capability boundaries in tabular form. We view the evaluation framework for Generation~4 TTS as an open research problem in its own right --- one that must jointly address coherence, expressiveness, instruction fidelity, scene realism, and interaction naturalness. We plan to tackle this in future work. We encourage readers to experience the system's expressiveness through direct listening; demo links can be found via web search.

\section{Conclusion, Limitations, and Future Work}
\label{sec:conclusion}

Our system rests on three pillars: \textbf{Label instead of filter} --- reclaiming ``dirty'' data as controllable acoustic dimensions; \textbf{Understand before synthesize} --- using CoT to surface implicit prosody decisions as explicit, editable reasoning; and \textbf{Hierarchical annotation as structured semantic interface} --- enabling an LLM Agent's scene-level understanding to flow losslessly through the \textit{Global-Sentence-Token} protocol stack to the synthesis engine, bridging arbitrary input modalities to arbitrary speech forms.

\paragraph{Limitations.}
The system is currently optimized for \textbf{content-creation scenarios} --- podcasts, audiobooks, film narration, and similar offline production tasks --- where generation quality and expressiveness take priority over latency. The system has not yet been adapted for real-time interactive settings (voice conversations, live-streaming, etc.), which demand millisecond-level response times, streaming output, and the ability to adjust expression on the fly based on user feedback. Bringing this expressiveness to real-time interaction is a central goal of our next phase. Additionally, the training data remains overwhelmingly speech-centric; dedicated sound-effect and music corpora have not yet been incorporated. The sound-effect and music generation that the system currently exhibits is entirely emergent --- learned incidentally from background audio in podcasts, interviews, and sports commentary. While encouraging, this emergent capability has a clear ceiling relative to purpose-built training.

\paragraph{Future Work.}
Five directions stand out:

\textit{From Speech to Sound.}
The \textit{Global-Sentence-Token} schema is modality-agnostic: its natural-language captions can describe a mournful cello solo or the dense patter of rain on a tin roof just as easily as speech. Adding large-scale sound-effect and music data --- annotated under the same ``label, don't filter'' philosophy --- should unlock cross-modal generalization well beyond what background-audio emergence achieves today.

\textit{Reference-Speech Timbre Control.}
The system already supports natural-language timbre control (``mature female, deep and husky''), providing considerable creative freedom. Building on this foundation, we plan to add a reference-audio capability: users will be able to supply a short recording of a target speaker, and the system will lock onto that timbre while preserving instruction-based control over expression style --- striking an optimal balance between creative flexibility and voice consistency.

\textit{From Content Creation to Real-Time Interaction.}
The system already delivers strong expressiveness in content-creation workflows. The next step is to bring these capabilities into real-time interactive settings --- streaming synthesis, low-latency inference, and dynamic expression adjustment driven by live user feedback. This will require an end-to-end streaming architecture that preserves high expressiveness, enabling it to not only \textit{produce} content but also \textit{perform live} in conversations and broadcasts.

\textit{RL and Inference Efficiency.}
Training on a broad clean-to-noisy, simple-to-complex distribution already reduces hallucination; we hypothesize it will also yield more informative reward signals for RL. Validating this, along with optimizing inference speed for practical deployment, is a priority.

\textit{Evaluation Methodology for Expressive Long-Form Speech.}
The absence of a reliable automated evaluation framework for expressive long-form speech synthesis remains a critical gap. Existing metrics --- DNSMOS, PESQ, CLAP, and Audio Captioning models --- address signal quality or short-clip semantics, but none can adequately assess emotional-arc coherence, multi-speaker interaction naturalness, scene fidelity, or instruction-following accuracy at the paragraph or document level. We plan to develop a dedicated multi-dimensional evaluation suite, drawing on LLM-as-judge paradigms for semantic and prosodic alignment as well as structured human perceptual studies that probe specific expressiveness dimensions. Establishing such a benchmark is, in our view, a prerequisite for systematic progress toward Generation~4 TTS.

\bibliographystyle{unsrtnat}
\bibliography{any2speech}

\end{document}